\documentclass[twocolumn,aps,pra,showpacs]{revtex4}
\usepackage{graphicx}
\usepackage{amsmath}
\usepackage{hyperref}


\begin{document}

\title{Measurement of the forbidden tensor polarizability of Cs using an
all-optical Ramsey resonance technique}

\author{C. Ospelkaus}
\altaffiliation[Present address:]{ Institut f\"ur Laser-Physik,
Jungiusstrasse 9, D-20355 Hamburg, Germany }

\author{U. Rasbach}
\altaffiliation[Present address:]{
 Institut f\"ur Angewandte Physik, Wegelerstrasse 8,
D-53115 Bonn, Germany }

\author{A. Weis}

\email{antoine.weis@unifr.ch}

\homepage{http://www.unifr.ch/physics/frap/}

\affiliation{D\'epartment de Physique, Universit\'e de Fribourg,
1700~Fribourg / Switzerland}

\begin{abstract}
We have measured the strongly suppressed electric tensor polarizability of
the Cs ground state using an optical pump-probe technique in a thermal
atomic beam. The result $\alpha _{2}(F=4)/h=$-3.34(2)$_{stat.}$(25)$_{syst.}$%
10$^{\text{-8}}\,$Hz/(V/cm)$^{2}$ agrees with a previous measurement and
confirms the long-standing discrepancy with a theoretical value. The
anticipated future reduction of the total uncertainty to the 1\% level makes
this quantity a valuable test for atomic structure calculations involving
short-range inner-atomic interactions.
\end{abstract}

\pacs{32.60.+i, 12.20.Fv}

\maketitle

\section{introduction}

Precision measurements of atomic properties provide an important testing
ground for atomic structure calculations. The commonly investigated
lifetimes, (allowed) Stark shifts and transition oscillator strengths depend
on electric dipole matrix elements and their measurement provides
information mainly on the long-range parts of atomic wave functions. There
is, on the other hand, a strong interest in testing the short-range behavior
of atomic wave functions as it plays a fundamental role in the measurement
and calibration of parity-violating effects in atomic systems induced by the
short-range weak interactions. Hyperfine and spin-orbit coupling constants
are well-known examples of atomic properties, which dominantly depend on the
short-range properties of the atomic wave function. Forbidden tensor
polarizabilities provide complementary tests of short range interactions in
atoms.

The electric field-induced shift of a hyperfine Zeeman component $\left|
nL_{J},F,M_{F}\right\rangle $ is given by
\begin{equation}
\Delta E(nL_{J},F,M_{F})=-\frac{1}{2}\alpha (nL_{J},F,M_{F})\mathcal{E}^{2},
\label{EStark1}
\end{equation}
where the polarizability $\alpha $ has scalar ($\alpha _{0}$) and tensor ($%
\alpha _{2}$) parts
\begin{equation}
\alpha (nL_{J},F,M_{F})=\alpha _{0}(F)+\alpha _{2}(F)\frac{3M_{F}^{2}-F(F+1)%
}{F(2F-1)}.  \label{EStark2}
\end{equation}

In second-order perturbation theory a non-vanishing tensor polarizability
arises only for states $\left| nL_{J}\right\rangle $ with a total electronic
angular momentum $J\geq 1$, so that $\alpha _{2}$ vanishes in alkali ground
states $\left| nS_{1/2}\right\rangle $. The scalar polarizabilities of such
states depend on
\begin{equation*}
\frac{\left| D(nS_{1/2},n^{\prime }P_{J})\right| ^{2}}{\Delta
E(nS_{1/2},n^{\prime }P_{J})}\,,
\end{equation*}
where $D(nS_{1/2},n^{\prime }P_{J})$ are $S-P$ dipole matrix elements and $%
\Delta E$ the corresponding energy splittings. It was first pointed out by
Sandars \cite{San67,San68} that a finite, but strongly suppressed (seven
orders of magnitude in the case of Cs) value of the tensor polarizability
arises as a third-order perturbation when the hyperfine interaction is taken
into account, leading to terms
\begin{equation}
\frac{\left| D(nS_{1/2},n^{\prime }P_{J})\right| ^{2}E_{hf}(nS_{1/2})}{%
\Delta E^{2}(nS_{1/2},n^{\prime }P_{J})}  \label{thirdorderS}
\end{equation}
and
\begin{equation}
\frac{\left| D(nS_{1/2},n^{\prime }P_{J})\right| ^{2}E_{hf}(n^{\prime }P_{J})%
}{\Delta E^{2}(nS_{1/2},n^{\prime }P_{J})}\,,  \label{thirdorderP}
\end{equation}
where $E_{hf}$ are matrix elements of the hyperfine interaction. The first
term, involving the Fermi contact interaction, leads to a shift of the
hyperfine splitting, which has been measured on the Cs clock transitions
\cite{Za57,Sim98}. This effect is approximately two orders of magnitude
larger than the tensor effect discussed here (second term), which, for Cs
atoms in fields of a few 10 kV/cm, induces Zeeman sublevel splittings on the
order of 1 Hz. The expressions (\ref{thirdorderS},\ref{thirdorderP})
illustrate the dependence of $\alpha _{2}$ on both long-range and
short-range interactions.

Tensor polarizabilities of alkalis were measured in the 1960's using
conventional Ramsey resonance spectroscopy \cite{Gou69}, but yielded large
discrepancies with theoretical values \cite{San67,San68}. Here we apply a
novel Ramsey technique to measure the tensor Stark effect of cesium. Our
all-optical method does not use radio-frequency (r.f.) or microwave fields,
thereby avoiding r.f. power dependent systematic effects, observed in
earlier experiments(cf. \cite{Gou69}).\bigskip

\section{Faraday-Ramsey Spectroscopy}

\begin{figure}[tbp]
{\centering \resizebox*{1\columnwidth}{!}{\includegraphics{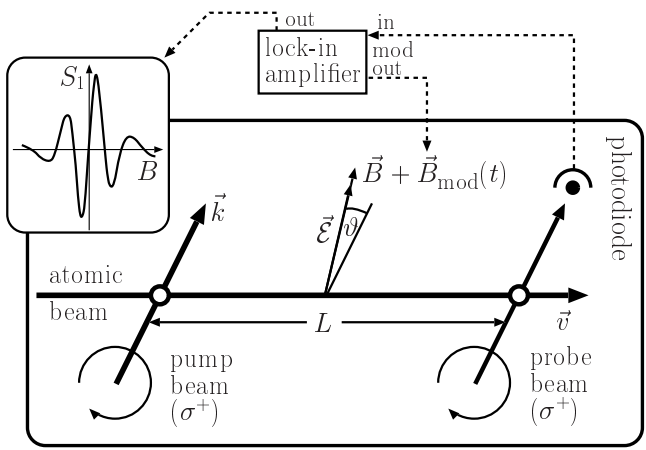}} }
\caption{Experimental setup of the pump-probe experiment with circularly
polarized light. The insert shows a typical experimental recording of the
probe beam absorption as a function of the magnetic field. With $L=30\,%
\mathrm{cm}$, the period of the oscillatory structure is approximately 200
nT.}
\label{fig_princip}
\end{figure}

The measurements presented in this paper were performed using an optical
pump-probe technique in a thermal cesium beam (figure \ref{fig_princip}). A
circularly polarized pump beam resonant with a hyperfine component of the Cs
$\mathrm{D}_{2}$ line (852 nm) produces spin polarization $\vec{S}=\langle
\vec{F}\rangle $ oriented along the $\vec{k}$-vector of the pump beam and
perpendicular to the atomic velocity $\vec{v}$. When exposed to a static
magnetic field $\vec{B}$ oriented at an angle $\vartheta $ with respect 
to $\vec{k}$, the spin polarization precesses around the magnetic field 
at the Larmor frequency $\omega _{B}=\gamma_{F} \cdot |\vec{B}|$.

The spin polarization is measured downstream by recording the transmission
of a weak circularly polarized probe laser beam. The absorption depends on
the projection of the velocity averaged spin polarization on the probe laser
beam. The magnetic field dependence of the transmitted intensity then leads
to a damped oscillatory pattern (Ramsey fringes) symmetric around $B=0$. For
practical purposes it is more convenient to work with a dispersive
(antisymmetric around $B=0$) fringe pattern. Such a line shape can be
generated as the derivative of the absorptive fringe pattern by applying 
a small amplitude modulation to the magnetic field and
detecting the transmitted intensity using a lock-in amplifier locked to the
modulation frequency. In additition to the well-known advantages of 
phase-sensitive detection, the lock-in amplifier will eliminate the 
$B$-independent parts of the signal which would otherwise make the pattern 
almost unobservable. For $\vartheta =90^{\circ }$ the resulting lineshape $%
S_{1}(B)$ is given by
\begin{equation*}
S_{1}(B)\propto \int_{0}^{\infty }\sin (\gamma _{F}B\frac{L}{v})\frac{\rho
(v)}{v}dv,
\end{equation*}
where $\rho (v)\,dv$ is the velocity distribution of the atomic density in
the probe region. A typical experimental recording of such a Ramsey fringe
pattern is shown as insert in figure \ref{fig_princip}. The steep central
zero-crossing of this pattern is a very sensitive discriminant for the
measurement of other (e.g. $\vec{\mathcal{E}}$-field induced) interactions
which produce differential phase shifts of the magnetic sublevels of the
ground state. Previous experiments using this technique were carried out
with linearly polarized laser beams and detected the (Faraday) rotation of
the probe beam polarization. This technique is closely related to the
(single beam) nonlinear Faraday effect, which can be interpreted as a
three-step process \cite{KWW93,KWWH93}. This resemblance, together with
Ramsey's idea of separated fields gave this method the name of
Faraday-Ramsey spectroscopy (FRS) \cite{WSKH93,SWKH93}. FRS has been used to
measure the Aharonov-Casher phase shift of $^{85}\mathrm{Rb}$ \cite{Goe95}.

The key idea of the experimental technique used to measure the tensor
polarizability is the following: In the interaction zone the Cs beam is
exposed to parallel static electric $\vec{\mathcal{E}}$ and magnetic $\vec{B}
$ fields oriented at an angle $\vartheta $ with respect to the $\vec{k}$%
-vector of the pump beam. When $\vec{\mathcal{E}}=0$ a feed-back loop
actively stabilizes the magnetic field to the center of the dispersive
central Ramsey fringe. Any additional phase shift induced by the electric
field is then compensated by an automatic adjustment of the magnetic field.
The corresponding compensation current $I_{\mathrm{FB}}$ in the field
generating coils is the signal of interest.

For the quantitative interpretation of this compensation technique one has
to know the electric field equivalent of the magnetic compensation field,
which can be calculated in a straightforward way by means of the Schr\"{o}%
dinger equation. A stretched hyperfine state $\left| F,M_{F}=F\right\rangle $
prepared in the pump region is allowed to evolve in the electric and
magnetic fields. For a monochromatic beam of velocity $v$ one then obtains
the wave function in the probe region, which allows to infer the $E,B$-field
induced change of the probe absorption coefficient:
\begin{equation}
S_{1}(\Phi _{B},\Phi _{E})\propto \Phi _{B}\sin ^{2}\vartheta +(2F-1)\Phi _{%
\mathcal{E}}\cos \vartheta \sin ^{2}\vartheta ,  \label{SmallFieldSig}
\end{equation}
where the magnetic and electric phases $\Phi _{B,\mathcal{E}}$ are given by
\begin{equation*}
\Phi _{B}=\gamma _{F}\frac{\int B\,dl}{v}\quad \mathrm{and}\quad \Phi _{%
\mathcal{E}}=-\frac{3}{2F(2F-1)}\frac{\alpha _{\mathrm{2}}}{\hbar }\frac{%
\int \mathcal{E}^{2}\,dl}{v}.
\end{equation*}
The integrals are over the atomic trajectories between the pump and probe
regions. As both phases have the same dependence on the atomic velocity $v$
any velocity averaging will produce a mere proportionality factor, which is
irrelevant for the following discussion, and will thus be omitted. Equation
\ref{SmallFieldSig} - derived under the assumption $\Phi _{B,\mathcal{E}}\ll
1$ - allows to define the optimal geometry and to calibrate the
electrically-induced shift in terms of the magnetic compensation. As the
feedback loop in the experiment adjusts the magnetic field such that $%
S_{1}(\Phi _{B},\Phi _{E})=0$, the signal $S$ is most sensitive to $\Phi _{%
\mathcal{E}}$, i.e. to the tensor polarizability, when $\vartheta $ is
chosen to be the magic angle $\vartheta _{m}$, which satisfies $3\cos
^{2}\vartheta _{m}-1=0$ and which maximizes the electric contribution in
Equation \ref{SmallFieldSig}. The same equation can also be used to define
the magnetic equivalent of the electric phase shift, which implies
\begin{eqnarray}
\frac{\alpha _{2}}{h} &=&\frac{F\cdot \gamma _{F}}{3\pi \cos \vartheta }%
\cdot \frac{\int B\,dl}{\int \mathcal{E}^{2}\,dl}  \label{alfa2cal1} \\
&\equiv &\frac{F\cdot \gamma _{F}}{3\pi \cos \vartheta }\cdot \frac{\beta }{%
\varepsilon }\cdot \frac{I_{\mathrm{FB}}}{U_{\mathrm{HV}}^{2}}\,.
\label{alfa2cal2}
\end{eqnarray}
Equation \ref{alfa2cal2} can be understood as follows: the first factor
depends on specific atomic properties and on the geometry of the
experimental arrangement, while the second and third factors contain
calibration constants $\beta $ and $\varepsilon $, which relate the field
integrals $\int B\,dl$ and $\int \mathcal{E}^{2}\,dl$ to the feedback
current $I_{\mathrm{FB}}$ and to the square of the applied high voltage $U_{%
\mathrm{HV}}$ respectively. Once the calibration constants are known, the
tensor polarizability can be inferred from the measured dependence of $I_{%
\mathrm{FB}}$ on $U_{\mathrm{HV}}$.

\section{Experimental Setup}

In the experiment, the pump and probe beams are delivered by the same
extended cavity diode laser locked to the $F=4\rightarrow F^{\prime }=5$
hyperfine component of the Cs $\mathrm{D}_{2}$ transition using standard
saturated absorption spectroscopy in an auxiliary vapor cell. The laser beam
is transferred to the experiment proper by an optical fiber which also
serves as a mode cleaner. The output intensity of the fiber is actively
stabilized by a feed-back circuit controlling the input intensity with an
acousto-optic modulator. The atomic beam is produced by a reflux oven \cite
{SE81} operated at a temperature of $130^{\circ }\mathrm{C}$ and delivering
a beam with a divergence of 40 mrad. Between the pump and probe zones --
separated by a 30 cm long interaction zone -- the atomic beam propagates in
a 7 cm diameter electrically grounded tube, which contains a pair of
polished copper electrodes that can be rotated around the atomic beam axis.
One of the electrodes is grounded, while the other electrode is connected to
a computer-controlled high voltage power supply that allows to apply
electric fields up to 20 kV/cm. Two grounded diaphragms are located between
the electrodes and the laser beams to ensure that the pump and probe
interactions take place in an electric field-free environment. A solenoid
wound on the beam tube and two pairs of rectangular coils allow to apply
magnetic fields of arbitrary orientation and/or to shield unwanted field
components. All coils extend over the pump and probe regions. The whole
set-up is enclosed in a double cylindrical mu-metal shield (transverse /
longitudinal shielding factor 14000 resp. 5800). The data
acquisition (high voltage and feedback current) is controlled by a PC and
the robustness of the laser frequency lock allows the experiment to be run
overnight without user intervention.

\section{Systematic studies and results}

When $\vartheta \neq 90^{\circ}$ the motional magnetic field seen by the
atoms moving through the static electric field leads to an additional
magnetic phase shift, which is proportional to the electric field, i.e. to $%
U_{\mathrm{HV}}$. This linear Stark effect is also known as Aharonov-Casher
phase shift \cite{Goe95}. A further noisy background may come from slowly
drifting magnetic offset fields. The measurements were therefore performed
in a way that allows to eliminate these backgrounds from the recorded data.

In a typical experimental run (12 hours) the data acquisition software
controls the high voltage power supply by applying 5 discrete voltages $U_i
\,(i=1,\ldots,5)$ to the electrodes, each with both polarities. We start
with $U_{\mathrm{HV}}=0$ and record $I_{\mathrm{FB}}$ for two minutes. Then
the software applies the voltage $U_{\mathrm{HV}}=U_{i=1}$ and records $I_{%
\mathrm{FB}}$ for another two minutes, after which the polarity is reversed
and $I_{\mathrm{FB}}$ is recorded with $U_{\mathrm{HV}}=-U_{i=1}$ . In the
next step $i$ is incremented by 1 and the cycle starts again with $U_{%
\mathrm{HV}}=0$. When $i=5$ is reached, the whole procedure is repeated. The
integration time of 2 minutes was chosen after a detailed study of the
system stability in terms of the Allan variance of $I_{\mathrm{FB}} $, which
showed a minimal value for integration times between 100 and 200 seconds.

\begin{figure}[tbp]
{\centering
\resizebox*{0.95\columnwidth}{!}{\includegraphics{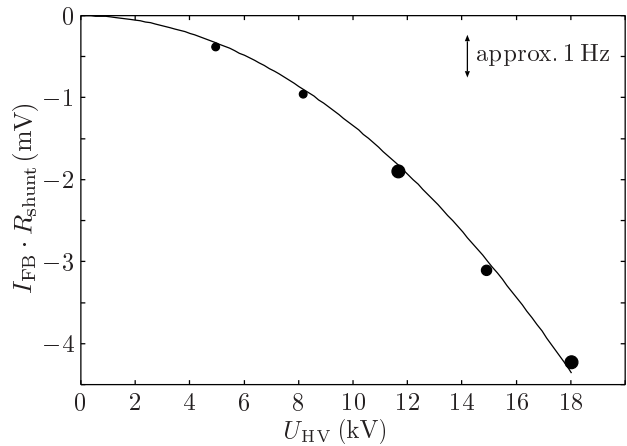}} }
\caption{Quadratic Stark shift due to tensor polarizability after
substraction of an offset and motional field contributions. The 
vertical statistical error bar of each data point corresponds to 
the vertical dot size. The horizontal error is negligible. }
\label{fig_parabel}
\end{figure}

In the off-line data analysis the average value of $I_{\mathrm{FB}}$ for
each cycle of a given high voltage is computed. Then the difference of time
consecutive measurements with and without high voltage are calculated in
order to subtract base line drifts and finally the values thus obtained for
consecutive measurements with reversed polarities are averaged in order to
eliminate the contribution from the motional field effect. The resulting
averaged values then show a pure quadratic dependence on $U_{\mathrm{HV}}$
as illustrated in figure \ref{fig_parabel}.

A simple quadratic fit to the data points allows, together with the
calibration constants $\beta $ and $\varepsilon $, to infer $\alpha_{2} $
using equation \ref{alfa2cal2} with a typical statistical error of 1\% per
run. The magnetic calibration constant $\beta $ was inferred from the Ramsey
fringe pattern with a precision of 0.3\%, while the electric calibration
constant $\varepsilon $ was determined by a semi-emprical method involving
both numerical calculations using an adaptive boundary element method and
the Aharonov--Casher phase shift (precision of 0.3\%). Details of these
calibration procedures will be published elsewhere \cite{FCal}.

The automated measurements have allowed us to study different systematic
effects. No significant dependence of the results on doubling or halving the
pump laser intensity was found. This proves that the beam is fully polarized
in the $\left|F=4,M_{F}=4\right>$ hyperfine state, which is a prerequisite
for the validity of equation \ref{SmallFieldSig}. There was also no
detectable dependence of the results on slight misalignments of the
quarter-wave plates producing the circularly polarized light, nor on
variations of the oven temperature and hence the velocity distribution of
the beam. Variations of the probe intensity yielded no effect on the
obtained results. There is however a serious systematic uncertainty, which
currently limits the absolute precision of the results. Equations \ref
{alfa2cal1} and \ref{alfa2cal2} were derived by assuming that the magnetic
and electric fields are perfectly parallel. The orientation of the electric
field was realized by a mechanical rotation of the electrode structure and
could be determined with an accuracy of better than 8.7 mrad. For historical
reasons the rotated magnetic field was realized by adding the fields of two
orthogonal pairs of coils. The contribution from the uncertainty (44 mrad)
of this alignment to the final result was estimated from a
systematic experimental study. The uncertainties for $\varepsilon $, $\beta $
and $\vartheta $, together with the dominating error from the relative
orientation of the $\mathcal{E}$ and $B$ fields yield a (conservative) upper
bound for the total systematic uncertainty of $7.5\%$.

After averaging five runs the overall statistical error is 0.7\%, which
leads to the preliminary result of
\begin{equation*}
\alpha _{\mathrm{2}}(F=4)/h=-3.34(2)(25)\cdot 10^{-8}\frac{\mathrm{Hz}}{(%
\mathrm{V}/\mathrm{cm})^{2}},
\end{equation*}
where the numbers in parentheses give the statistical and systematic errors
respectively. The value of $\alpha _{2}(F=4)$ inferred from previously
reported \cite{Gou69} experimental results is
\begin{equation*}
\alpha _{\mathrm{2}}(F=4)/h=-3.66(21)(7.3)\cdot 10^{-8}\frac{\mathrm{Hz}}{(%
\mathrm{V}/\mathrm{cm})^{2}}\,
\end{equation*}
while the theoretical prediction from \cite{San68,San67} quoted in the
article by Gould et al. \footnote{%
according to Equations \ref{EStark1} and \ref{EStark2} the shift rate $%
\Delta \nu / \mathcal{E}^{2}$ reported in \cite{Gou69} has to be multiplied
by 8/3 in order to infer $\alpha_{2}(F=4)$.} is
\begin{equation*}
\alpha _{\mathrm{2}}(F=4)/h=-4.133\cdot 10^{-8}\frac{\mathrm{Hz}}{(\mathrm{V}%
/\mathrm{cm})^{2}}.
\end{equation*}

\section{Conclusion and Outlook}

We have demonstrated a novel technique for the measurement of tensor
polarizabilities in alkali atoms. We have shown that with $^{133}\mathrm{Cs}
$ a statistical precision below the 1\% level can be achieved. The technique
can easily be extended to other atoms by using suitable atomic beams and
light fields. Our result is consistent with an earlier experimental result,
and confirms the previously reported discrepancy with theoretical
calculations at our current level of precision. In order to overcome the
main current source of systematic error (parallelism of $\vec{\mathcal{E}} $
and $\vec{B} $), a new precision electrode arrangement with integrated
magnetic field coils is currently under construction. With this improved
set-up, a measurement of $\alpha _{\mathrm{2}} $ at the 1\% level will be
within reach.

\begin{acknowledgments}
The authors thank the mechanical workshop at the Physics
Department of the University of Fribourg for their skillful
support. The authors thank the mechanical workshop and J.-L. Schenker for 
skillful support. We acknowledge the contributions of B. Schuh, M. Niering, 
and F. Rex in early stages of this project. This work was financed in parts 
by Schweizerischer Nationalfonds (SNF) and Deutsche
Forschungsgemeinschaft (DFG).
\end{acknowledgments}

\bibliographystyle{apsrev}
\bibliography{tenspol}

\end{document}